\documentclass[notitlepage, aps, prb, twocolumn,  superscriptaddress, amsmath,amssymb,10pt]{revtex4-1}

\usepackage{graphicx}         
\usepackage{hyperref}
\hypersetup{
	linkcolor=blue,
	citecolor=blue,
	filecolor=blue,
	urlcolor=blue,
	colorlinks=true
}

\usepackage[usenames,dvipsnames]{color}
\usepackage{soul} 
\soulregister\cite7
\soulregister\ref7
\soulregister\eqref7
\usepackage{comment}

\newcommand{\tr}{\ensuremath{\text{tr}}}
\renewcommand{\v}[1]{\ensuremath{\boldsymbol{\mathbf{#1}}}} 
\newcommand{\abs}[1]{\left| #1 \right|}

\newcommand{\UIUCECE}[0]{Department of Electrical and Computer Engineering, University of Illinois at Urbana-Champaign, Urbana, IL 61801, USA}
\newcommand{\UIUCPhys}[0]{Department of Physics, University of Illinois at Urbana-Champaign, Urbana, IL 61801, USA}
\newcommand{\UIUCMNTL}[0]{Micro and Nanotechnology Laboratory, University of Illinois, 208 N. Wright Street, Urbana IL 61801, USA}
\newcommand{\Stanford}[0]{Department of Electrical Engineering, Stanford University, Stanford, California 94305, USA}

\makeatletter
\newcommand*{\citenump}[2][]{%
  \begingroup
  \let\NAT@mbox=\mbox
  \let\@cite\NAT@citenum
  \let\NAT@space\NAT@spacechar
  \let\NAT@super@kern\relax
  \renewcommand\NAT@open{[}%
  \renewcommand\NAT@close{]}%
  \citep[#1]{#2}%
  \endgroup
}
\makeatother

\allowdisplaybreaks 

\begin{abstract}
	 Recent work has extended topological band theory to open, non-Hermitian Hamiltonians, yet little is understood about how non-Hermiticity alters the topological quantization of associated observables. We address this problem by studying the quantum anomalous Hall effect (QAHE) generated in the Dirac surface states of a 3D time-reversal-invariant topological insulator (TI) that is proximity-coupled to a metallic ferromagnet. By constructing a contact self-energy for the ferromagnet, we show that in addition to generating a mass gap in the surface spectrum, the ferromagnet can introduce a non-Hermitian broadening term, which can obscure the mass gap in the spectral function. We calculate the Hall conductivity for the effective non-Hermitian Hamiltonian describing the heterostructure and show that it is no longer quantized despite being classified as a Chern insulator based on non-Hermitian topological band theory. Our results indicate that the QAHE will be challenging to experimentally observe in ferromagnet-TI heterostructures due to the finite lifetime of quasi-particles at the interface.
\end{abstract}

\begin{document}

\title{Loss of Hall Conductivity Quantization in a\\Non-Hermitian Quantum Anomalous Hall Insulator}
\author{Timothy M. Philip}\email{tphilip3@illinois.edu}\affiliation{\UIUCECE}\affiliation{\UIUCMNTL}
\author{Mark R. Hirsbrunner}\affiliation{\UIUCMNTL}\affiliation{\UIUCPhys}
\author{Matthew J. Gilbert}\affiliation{\UIUCECE}\affiliation{\UIUCMNTL}\affiliation{\Stanford}
\maketitle





%
%


\section{Introduction}

The last decade has seen a revolution in the understanding of the electronic structure of solids with the formulation and development of topological band theory, which provides a unified system to classify materials ranging from insulators and semi-metals to superconductors using topological invariants.\cite{Hasan2010,Qi2011,Bernevig2013} These quantized topological invariants provide a robust classification for materials, as they cannot be changed by adiabatic deformations of the systems. An important consequence of a non-trivial topological classification is that some response of the system to an external stimulus is also quantized proportional to its topological invariant. One well-known example of this quantization is in the integer quantum hall effect (IQHE) or the quantum anomalous Hall effect (QAHE), in which the Hall conductivity is given as 
\begin{equation}
\sigma_{yx} = -\nu_\text{occ} \frac{e^2}{h}, \label{eq:quantized_sigmaxy}
\end{equation}
where $e$ is the electron charge, $h$ is Planck's constant, and $\nu_\text{occ}$ is the sum of the TKNN invariants or Chern numbers of occupied bands.\cite{Klitzing1980a,Haldane1988} Because of the topological quantization of $\sigma_{yx}$, the Hall response is remarkably robust to perturbations and the presence of disorder, allowing for experimental measurements of the IQHE accurate to a few parts in $10^{10}$ of the theoretically-predicted, quantized value.\cite{Jeckelmann1997a} Despite this success in predicting the quantization of the Hall conductivity, topological band theory is formulated for closed, Hermitian Hamiltonians, and it is, therefore, unclear if and how open systems can be topologically classified.

To address this issue, recent studies have extended topological band theory to characterize non-Hermitian Hamiltonians,\cite{San-Jose2016,Lieu2018,Cerjan2018,Zyuzin2018,Yao2018,Zhou2018a,Shen2018a} which arise in systems that are opened to  external reservoirs or interactions with other particles. Notably, non-Hermitian Hamiltonians can host topological phases and invariants that cannot be seen in Hermitian systems, resulting in unusual predictions such as bulk Fermi arcs in 2D systems.\cite{Papaj2018,Zhou2018a} Despite this progress in the understanding of non-Hermitian systems and their topological classifications, the effect of non-Hermiticity on the quantization of physical observables is not well-understood. 

In this work, we explore the consequences of non-Hermiticity on physical observables by quantitatively studying the QAHE generated in the Dirac surface states of a 3D time-reversal-invariant topological insulator (TI) when proximity-coupled to a metallic ferromagnet. In addition to a time-reversal-breaking mass gap generated in the Dirac surface spectrum by the ferromagnet, we see that the presence of metallic bands at the Dirac point give the surface states a finite lifetime, as electrons can escape into the ferromagnet. This finite lifetime results in broadening of the states that is comparable in magnitude to the mass gap, which in turn, results in a gapless spectral function. To characterize the impact of the this broadening on the QAHE, we calculate the Hall conductivity of this system via the Kubo-Streda formula. 
Non-Hermitian topological band theory suggests that such a system retains its classification as a Chern insulator, but we find that the Hall conductivity is no longer quantized as in Eq.~\eqref{eq:quantized_sigmaxy}. 
We compare the proximity-coupled case to one where the mass gap is generated by bulk magnetic dopants and find that the broadening due to magnetic impurity scattering is much smaller than the mass gap, thus allowing for the observation of the QAHE in these systems. Our results show that the non-Hermiticity introduced in open topological systems causes the loss of topological quantization of observables   and can severely limit the ability to experimentally observe such responses.

\begin{figure*} 
	\includegraphics[width=\textwidth]{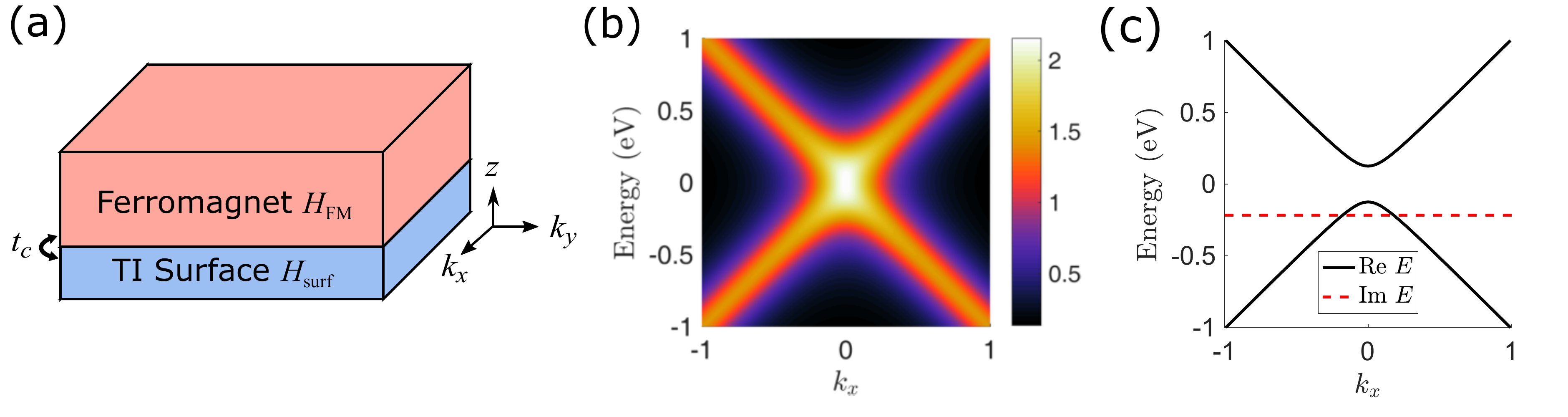}
	\caption{(a) Schematic of the proximity-coupled heterostructure. The topological insulator surface, given by the Hamiltonian $H_\text{surf}$, is proximity-coupled to a semi-infinite ferromagnet, given by the Hamiltonian $H_\text{FM}$, where $t_c$ is the strength of the coupling. (b) The spectral function of the effective Hamiltonian, given by Eq.~\eqref{eq:H_eff}, that describes the proximity-coupled heterostructure with parameters $\alpha = 1$, $M_\text{FM} = 1$, $t_0 = 1$, and $t_c = 0.5$. Although the ferromagnet generates a time-reversal breaking mass gap in the surface states, the broadening that is also introduced by the presence of metallic bands at the Dirac point is large enough to result in a gapless spectrum. (c) The eigenvalues for the Hamiltonian with the same parameters reveals that the real part of the spectrum is gapped by the mass term and the non-Hermitian contribution simply shifts the eigenvalues into the complex plane. Since the bands remain separable, the system retains its classification as a Chern insulator.}\label{fig:spectral_function}
\end{figure*}

\section{Model Hamiltonian and Ferromagnet Contact Self-Energy}

Figure~\ref{fig:spectral_function}(a) depicts a schematic for the TI-ferromagnet heterostructure we study. The low-energy, effective Hamiltonian for the surface states of a 3D time-reversal-invariant TI is given by the 2D Dirac equation
\begin{equation}
	H_\text{surf}(\v k) = \alpha (k_y  \sigma_x - k_x \sigma_y), \label{eq:TI_Surface_Hamiltonian}
\end{equation}
where $\v k = (k_x, k_y)$ is the momentum of the electron,  $\alpha = \hbar v_F$, $v_F$ is the Fermi velocity of the surface electrons, and $\sigma_i$ are the spin Pauli matrices. We model the ferromagnet with a tight-binding Hamiltonian with nearest-neighbor hopping in the $\v{ \hat z}$ direction given by
\begin{align}
	\begin{split}
		H_\text{FM}(\v k) = \sum_{ z} &\left[  \psi_{\v k, z}^\dag H_\text{on}(\v k, z) \psi_{\v k, z}  \right. \\
			&\hspace{2em}\left.+  \left ( \psi_{\v k, z}^\dag H_\text{hop} \psi_{\v k, z + a \hat{\v z}}  + \text{H.c.}\right)\right],
	\end{split} \label{eq:FM_Metal_Hamiltonian}
\end{align}
where $\psi_{\v k, z}^{\dag}\,(\psi_{\v k, z})$ is the creation (annihilation) operator for an electron with in-plane momentum $\v k = (k_x, k_y)$ and position $z$, $H_\text{on} = \epsilon(\v k) \,\sigma_0 + M_\text{FM} \,\sigma_z$ is the on-site term, $\epsilon(\v k)$ is the in-plane dispersion of the metallic bands of the ferromagnet, $M_\text{FM}$ is the spin-splitting energy within the ferromagnet, and $H_\text{hop} = - t_0 \,\sigma_0$ is the hopping matrix in the $\v{\hat z}$ direction. To understand the effect of proximity coupling a ferromagnet to a topological insulator (TI) surface state, we calculate the contact self-energy that fully captures the effect of a semi-infinite ferromagnetic contact. To obtain the self-energy, we first must compute the surface Green function, $g(E)$, of the contact, which for a semi-infinite, uniform material follows the equation\cite{Pourfath2014} 
\begin{equation}
	[A(E) - H_\text{hop} g(E) H_\text{hop}^\dag] g(E) = I, \label{eq:surface_gr_recursive}
\end{equation}
where $A(E) = EI - H_\text{on}$, $E$ is the energy of interest, $I$ is the identity matrix, $H_\text{on}$ is the on-site Hamiltonian matrix for the surface of the contact, and $H_\text{hop}$ is the hopping matrix perpendicular to the contacting surface. For general Hamiltonians, the solution to Eq.~\eqref{eq:surface_gr_recursive} is non-analytic, but for Hamiltonians that have the specific property that $H_\text{on}$ and $H_\text{hop}$ are diagonal, as is the case for the ferromagnet Hamiltonian in Eq.~\eqref{eq:FM_Metal_Hamiltonian}, an analytic closed-form solution can be obtained. Once the surface Green function of the contact is found, the contact self-energies are given simply as
\begin{align}
	\Sigma_c(E) &= H_\text{coupling}^\dag g(E) H_\text{coupling}, \label{eq:contact_Sigma}
\end{align}
where $H_\text{coupling}$ is the coupling matrix between the system of interest and the contact. We assume that there is no spin mixing at the interface resulting in a diagonal matrix $H_\text{coupling} = -t_c \sigma_0$. By solving Eqs.~\eqref{eq:surface_gr_recursive} and \eqref{eq:contact_Sigma}, we obtain a contact self-energy for the ferromagnet with the form
\begin{equation}
	\Sigma_c(E) = 
	\begin{bmatrix}
		\Sigma_{\uparrow} & 0 \\
		0 & \Sigma_{\downarrow}
	\end{bmatrix},  \label{eq:SigmaFM}
\end{equation}
where the diagonal components are given as
\begin{widetext}
\begin{equation}
	\Sigma_{\uparrow/\downarrow} = \begin{cases}
						\frac{\abs{t_c}^2}{2\abs{t_0}^2} \left(E - \epsilon(\v k) \mp M_\text{FM} + \sqrt{ ( E - \epsilon(\v k) \mp M_\text{FM} )^2 - 4\abs{t_0}^2} \right)  &  E \leq  \epsilon(\v k) + M_\text{FM},  \\
						\frac{\abs{t_c}^2}{2\abs{t_0}^2} \left( E - \epsilon(\v k) \mp M_\text{FM} - \sqrt{ ( E - \epsilon(\v k) \mp M_\text{FM} )^2 - 4\abs{t_0}^2} \right)  & E >  \epsilon(\v k) + M_\text{FM},
					\end{cases}  \label{eq:SigmaFM_down} 
\end{equation}
\end{widetext}
and the upper (lower) sign corresponds to up (down) spin (See Supplemental Material Sec.~I for a detailed derivation).


\section{Non-Hermitian Effective Hamiltonian}

To understand the impact of metallic bands at the Dirac point, we focus on the case where the bands of the ferromagnet are centered around $E=0$ such that, at low momenta, $\epsilon(\v k) \to 0$. In addition, we impose the constraint $M_\text{FM} < 2 t_0$ to ensure that the spin up and spin down bands do not completely separate in energy to create a ferromagnetic insulator. Within this regime, we can understand the influence of the metallic ferromagnet by studying the low-energy limit of the contact self-energy in Eq.~\eqref{eq:SigmaFM}-\eqref{eq:SigmaFM_down}:
\begin{equation}
 	\Sigma_c \approx -i\frac{\abs{t_c}^2}{2\abs{t_0}^2} \sqrt{ 4\abs{t_0}^2 - M_\text{FM}^2} \,\sigma_0 - \frac{\abs{t_c}^2 }{2\abs{t_0}^2}M_\text{FM} \,\sigma_z.
\end{equation}
Utilizing this approximation, we create an effective Hamiltonian that describes the TI surface states in the presence of a proximity-coupled ferromagnet as
\begin{align}
	\begin{split}
	H_\text{eff} &= H_\text{surf} + \Sigma_c \\
		&=   -i \Gamma \sigma_0 + \alpha (k_y \sigma_x  - k_x \sigma_y ) - M  \sigma_z , 
	\end{split}\label{eq:H_eff}
\end{align}
where $\Gamma \equiv \frac{\abs{t_c}^2}{2\abs{t_0}^2} \sqrt{ 4\abs{t_0}^2 - M_\text{FM}^2} $ and $M \equiv  \frac{\abs{t_c}^2 }{2\abs{t_0}^2}M_\text{FM}$.  As expected, the proximity-coupled ferromagnet introduces a time-reversal breaking term proportional to the exchange interaction strength in the ferromagnet. In addition, the presence of the metallic bands from the ferromagnet introduces a non-Hermitian broadening term that gives the surface states a finite-lifetime as surface state electrons can escape into the ferromagnet. Broadening is a common consequence of a integrating out the effect of an interaction or coupling to an external reservoir, but this self-energy is notable in that the broadening, $\Gamma$, can exceed the mass gap, $M$, when $\sqrt{2} t_0 > M_\text{FM}$. This implies that the spectral function, given as $A(\v k, E) = -2 \text{Im} (G^r(\v k, E) - G^a(\v k, E) )$, where $G^r(\v k, E) = [E\sigma_0 - H_\text{eff}(\v k)]^{-1}$ is the Green function of the system and $G^{a} = G^{r\dag}$, can be gapless, as is demonstrated in Fig.~\ref{fig:spectral_function}(a), despite the fact that a mass gap has been generated in the surface spectrum.

When we inspect the the energy eigenvalues, given as $\epsilon(\v k) = -i\Gamma \pm \sqrt{ M^2 + \alpha^2 |\v k|^2}$ and plotted with $k_y = 0$  in Fig.~\ref{fig:spectral_function}(b), we see that the real part is gapped and exactly that of a massive Dirac electron. The non-Hermitian broadening simply shifts these eigenvalues by $-i\Gamma$ but does not close the gap in the complex energy spectrum. Since the bands remain separable with non-zero $\Gamma$ and the eigenvectors are unchanged from the Hermitian Hamiltonian,  the bands of this non-Hermitian Hamiltonian are connected to those of the Hermitian Chern insulator with $\nu = \pm \frac{1}{2}$.\cite{Shen2018a} When the Fermi energy is within the mass gap of a Hermitian Chern insulator, we anticipate that the Hall conductivity should be quantized to be $\sigma_{yx} = -e^2/2h$ for positive values of $M_\text{FM}$ as in Eq.~\eqref{eq:quantized_sigmaxy}. Since the non-Hermitian broadening introduced by the ferromagnet contact self-energy can be large enough that the gap in the spectral function is closed, it is not immediately obvious if the Hall conductivity continues to be exactly quantized for non-Hermitian Chern insulators.

\begin{figure*} 
	\includegraphics[width=\textwidth]{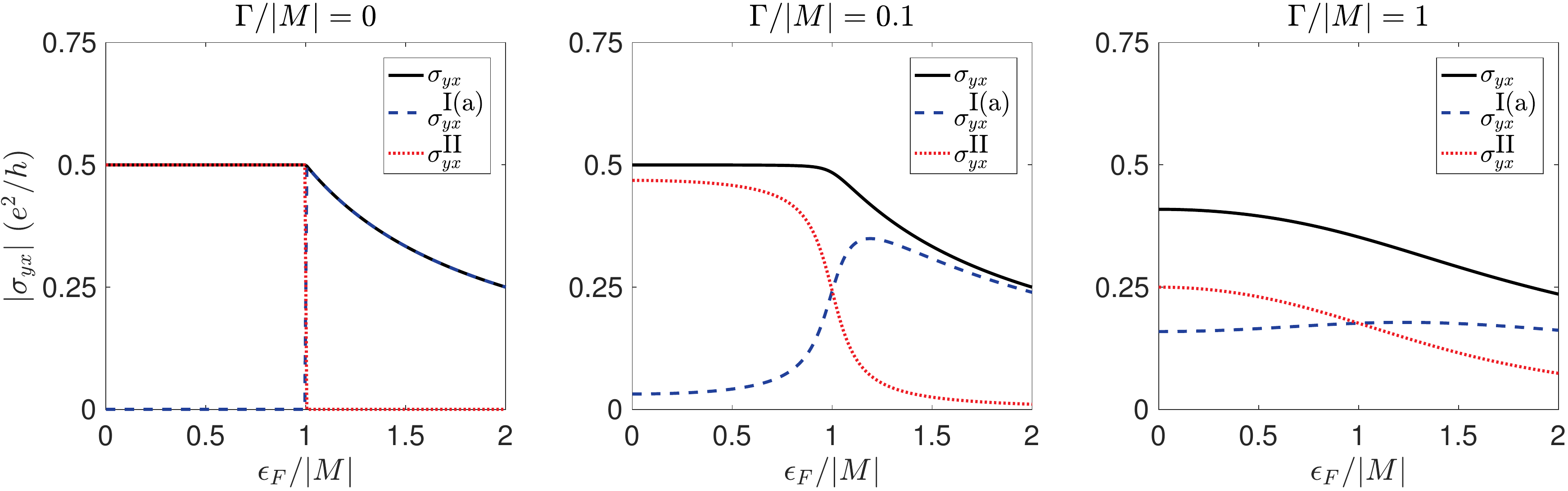}
	\caption{Hall conductivity of the effective Hamiltonian in Eq.~\eqref{eq:H_eff} as a function of the Fermi energy, $\epsilon_F$, for three values of $\Gamma/|M|$. When $\Gamma = 0$, the Hall conductivity is exactly quantized to $e^2/h$ when the Fermi energy is within the mass gap ($\epsilon_F/|M| < 1$). As the broadening,  $\Gamma$, is increased, we see that the total Hall conductivity monotonically decreases and is no longer exactly quantized.} \label{fig:sigmaxy}
\end{figure*}

\section{Hall Conductivity}

To quantify the impact of non-Hermiticity on the quantization of the topological observable in this system, we explicitly compute the DC Hall conductivity of the non-Hermitian Hamiltonian in Eq.~\eqref{eq:H_eff} using the Kubo-Streda formula,\cite{Streda1982a} which at zero temperature is given as
\begin{align}
	\sigma_{yx} &= \sigma_{yx}^\text{I(a)}  + \sigma_{yx}^\text{I(b)} + \sigma_{yx}^\text{II},
\end{align}
where 
\begin{align}
	\sigma_{yx}^{\text{I}(a)} &= \frac{e^2 \hbar}{2\pi V} \tr[   v_y G^r(\epsilon_F) v_x G^a(\epsilon_F) ],   \label{eq:sigmaxy_Ia}\\
	\begin{split}
		\sigma_{yx}^{\text{I}(b)} &= -\frac{e^2 \hbar}{4\pi V} \tr [  v_y G^r(\epsilon_F) v_x G^r(\epsilon_F) \\
		&\hspace{9em} + v_y G^a(\epsilon_F) v_x G^a(\epsilon_F) ], \label{eq:sigmaxy_Ib}
	\end{split}\\
	\begin{split}
		\sigma_{yx}^{\text{II}} &= \frac{e^2 \hbar}{4\pi V} \int_{-\infty}^{\epsilon_F} \!\! d\epsilon \, \tr \!\!\left[   
		  v_y G^r v_x \frac{ d G^r}{d \epsilon} 
		- v_y\frac{ d G^r}{d \epsilon}  v_x G^r \right. \\
		&\hspace{6em}\left. - v_y G^a v_x \frac{ d G^a}{d \epsilon} 
		+ v_y\frac{ d G^a}{d \epsilon}  v_x G^a
		\right] . 
	\end{split}\label{eq:sigmaxy_II}
\end{align}
Here  $V$ is the volume of space, $v_i$ is the velocity operator in the $\v{\hat i}$ direction, and $\epsilon_F$ is the Fermi energy. For the effective Hamiltonian in Eq.~\eqref{eq:H_eff}, the velocity operators are $v_x = -\frac{\alpha}{\hbar} \sigma_y$ and $v_y = \frac{\alpha}{\hbar} \sigma_x$.

The first two terms in this formulation, $\sigma_{yx}^{\text{I}(a)}$ and $\sigma_{yx}^{\text{I}(b)}$, are Fermi surface contributions and  are only non-zero when the Fermi energy crosses an energy band. The first term, $\sigma_{yx}^{\text{I}(a)}$, includes the intrinsic Berry phase component of the anomalous Hall conductivity in addition to extrinsic effects due to the presence of scattering mechanisms such as side-jump and skew scattering.\cite{Sinitsyn2007,Nunner2007a,Nagaosa2010a} The second term, $\sigma_{yx}^{\text{I}(b)}$, is identically zero for the Dirac Hamiltonian (See Supplemental Material Sec.~II and Ref.~\citenump{Nunner2007a}). The third term, $\sigma_{yx}^{\text{II}}$, corresponds to the contribution to the Hall conductivity that is due to the Fermi sea since the integration over energy can, in principle, include contributions from all occupied states. This contribution is quantized when the Fermi energy is within the mass gap and gives rise to the QAHE when a Chern insulating band is fully occupied. 



When $\Gamma$ is finite, the Fermi surface and Fermi sea contributions to the Hall conductivity take the closed form:
\begin{align}
	\sigma_{yx}^{\text{I}(a)}  &=  
		 - \frac{e^2}{h }\frac{M}{|\epsilon_F| 2\pi} \!\left[ \frac{\pi}{2} - \text{sgn}\, \epsilon_F \arctan\left( \frac{\Gamma^2 + M^2 - \epsilon_F^2 }{2 \Gamma \epsilon_F }  \right) \right]\!, \label{eq:sigmaxyIa_closed} \\
	\sigma_{yx}^{\text{II}}
		&= - \frac{e^2}{h}   \frac{   \text{sgn}\, M }{2\pi }\!\left[ \frac{\pi}{2} -  \arctan\left( \frac{ \Gamma^2 - M^2 + \epsilon_F^2 }{2 \Gamma |M|} \right) \right]\! \label{eq:sigmaxyII_closed}
\end{align}
(See Supplemental Material Sec.~II for a detailed derivation). Figure~\ref{fig:sigmaxy} shows the Hall conductivity for the effective non-Hermitian Hamiltonian as a function of Fermi level, $\epsilon_F$, at three different values of $\Gamma/|M|$. When $\Gamma = 0$ and when the Fermi energy is within the mass gap, the components of the Hall conductivity  take the expected form: $|\sigma_{yx}^{\text{I}(a)}|$ is identically zero, while $|\sigma_{yx}^{\text{II}}|$ is exactly quantized to $e^2/2h$.\cite{Sinitsyn2007,Nunner2007a} When $\Gamma$ is non-zero, however, both $\sigma_{yx}^{\text{I}(a)}$ and $\sigma_{yx}^{\text{II}}$ are non-zero within and above the mass gap. When $\Gamma/|M| \ll 1$, despite the the fact that $\sigma_{yx}^{\text{I}(a)} > 0$ and $\sigma_{yx}^{\text{II}} < e^2/2h$, the total Hall conductivity, $\sigma_{yx} = \sigma_{yx}^{\text{I}(a)} + \sigma_{yx}^{\text{II}}$, appears to remain quantized within the mass gap. 

The near quantization of the Hall conductivity for small broadening can be understood by expanding the expressions in Eqs.~\eqref{eq:sigmaxyIa_closed}-\eqref{eq:sigmaxyII_closed} in powers of $\Gamma/|M|$ (See Supplemental Material Sec.~III for more details). The leading order terms of the Fermi surface and sea contributions to the Hall conductivity are given as 
\begin{align}
	\sigma_{yx}^{\text{I(a)}} 
		\!&\approx \! -\frac{e^2}{h} \!  \left[\phantom{  \frac{   \text{sgn}\, M }{2} }+\frac{1}{\pi} \frac{ \Gamma M}{  M^2 - \epsilon_F^2 } - \frac{4M}{3\pi}  \frac{ \epsilon_F^2 \Gamma^3}{( M^2 - \epsilon_F^2)^3 } \right] \! \\
	\sigma_{yx}^{\text{II}}
		\!&\approx\! - \frac{e^2}{h}  \!  \left[  \frac{   \text{sgn}\, M }{2} - \frac{   1 }{\pi } \frac{ \Gamma M }{M^2 - \epsilon_F^2 } -  \frac{4M}{3\pi} \frac{\Gamma^3 M^2}{(\epsilon_F^2  - |M|^2)^3} \right]\!.
\end{align}
We see immediately that the terms that are first order in $\Gamma$ exactly cancel. Therefore, the leading-order correction to the quantized Hall conductivity within the mass gap is cubic in $\Gamma/|M|$:
\begin{align}
	\sigma_{yx} &\approx - \frac{e^2}{h} \left[ \frac{   \text{sgn}\, M }{2}  -  \frac{4\Gamma^3M}{3\pi} \frac{ M^2 + \epsilon_F^2}{(\epsilon_F^2  - |M|^2)^3}      \right].
\end{align}
Thus, when $\Gamma/|M| \ll 1$, the total Hall conductivity deviates negligibly from the quantized value.

When $\Gamma/|M|$ is comparable in magnitude to the mass gap, however, higher-order corrections are large enough to significantly decrease the total Hall conductivity from the quantized value. Thus, broadening can generate a distinct non-quantization of the Hall conductivity, in stark contrast to the robustness associated with Hermitian topological systems. In fact, we see from the expressions for the Hall conductivity in Eqs.~\eqref{eq:sigmaxyIa_closed}-\eqref{eq:sigmaxyII_closed} that any non-zero value of $\Gamma$ breaks the quantization of the Hall conductivity.

\subsection{Magnetic Proximity Effect}


\begin{table}
	\caption{Broadening and Induced Exchange Interaction for Common Ferromagnets}\label{table:estimation}
	\begin{tabular}{l || c | c | c | c | c | c  }
		&  $t_0\cite{Batallan1975}\,\text{(eV)}$ & $I\cite{Barreteau2016}\,\text{(eV)}$ & $m\cite{Cullity2009} \,(\mu_B)$ & $\Gamma\,\text{(eV)}$  & $M\,\text{(eV)}$ &  $\Gamma/|M|$ \\ \hline
	Fe  &     3.4     &   0.88   &      2.22     &     11.07      &    3.32  &     3.33    \\
	Ni  &     2.0     &   1.05   &      0.61     &      3.95      &    0.64  &     6.16    \\
	Co  &     2.6     &   1.10   &      1.76     &      6.27      &    2.52  &     2.49    
	\end{tabular}
\end{table}

Our characterization of the impact of non-Hermiticity on the quantization of the Hall conductivity in Chern insulating systems has clear ramifications for experimental observation of the QAHE in ferromagnetic-TI heterostructures. Because the non-Hermitian broadening introduced by the ferromagnet contact self-energy in Eqs.~\eqref{eq:SigmaFM}-\eqref{eq:SigmaFM_down}  is comparable to the mass gap in the magnetic proximity effect, the Hall conductivity deviates significantly from the quantized value predicted from the topological classification of the bands. In Table~\ref{table:estimation}, we estimate the broadening, $\Gamma$, and induced exchange interaction, $M$, for common ferromagnetic metals. We assume the exchange interaction within is given as $M_\text{FM} = Im$,\cite{Barreteau2016} where $I$ is Stoner parameter of the material and $m$ is the spin magnetization of the atoms in units of the Bohr magneton, $\mu_B$. For simplicity, we assume that the interface coupling is given as $t_c = t_0$ for each material. We see that the ratio of broadening to induced exchange interaction, $\Gamma/|M|$, is well above unity for iron, nickel, and cobalt. As such, when these ferromagnetic metals are placed in proximity to the surface a TI, we expect the Hall conductivity to be deviate from the quantized value.

\begin{figure}
	\includegraphics[width=.8\columnwidth]{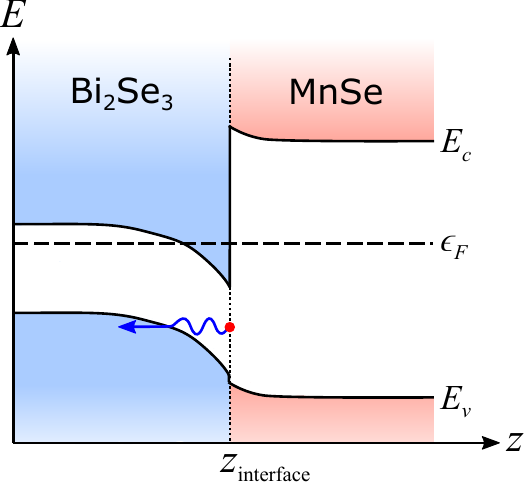}
	\caption{Band diagram of the junction between the TI Bi$_2$Se$_3$ and the ferromagnetic insulator MnSe, where $E_c$ is indicates the energy corresponding to the bottom of the conduction band of the materials and $E_v$ indicates the energy corresponding to the top of the valence band. Due to work function differences between the two materials, significant band bending can occur in the TI. Although the Dirac point, indicted by the red circle, lies within the band gap of MnSe, it is now below $E_v$ for $z < z_\text{interface}$, which allows surface state electrons to tunnel to the bulk valence band of Bi$_2$Se$_3$  and gives the surface states a finite lifetime. Adapted from Ref.~\citenump{Luo2013a}.}\label{fig:band_bending}
\end{figure}

Although our analysis has been limited to the specific case of a ferromagnetic metal with bands centered on the Dirac point, we note that large levels of broadening can arise even in heterostructures with ferromagnetic insulators when the Dirac point lies within the band gap of the ferromagnet. Figure~\ref{fig:band_bending} depicts the band diagram of a heterostructure of the TI Bi$_2$Se$_3$ and the ferromagnetic insulator MnSe.\cite{Luo2013a} Due to charge transfer generated by the work function difference between the two materials, a significant amount of band bending occurs at the interface. The Dirac point, indicated by the red circle, remains within the band gap of MnSe, which allows us to neglect broadening caused by the bulk bands of the ferromagnet. The bend bending on the TI side of the interface,  however, shifts the Dirac point below the top of the valence band of the TI. As such, the surface state electrons can tunnel through the potential barrier and escape into the bulk valence band of the TI, giving the surface states a finite lifetime similar to what we observe in metallic ferromagnets. Theoretical studies on band bending effects at the surface of TIs have already shown that significant broadening is generated through tunneling into bulk bands.\cite{Bahramy2012,Garate2012} Therefore, even when a ferromagnetic insulator is used to generate a mass gap in the TI surface states, the QAHE will be challenging to observe unless the work function difference between the materials is overcome using electrostatic gating.

\subsection{Magnetic Impurity Scattering}

We contrast our analysis of the mass gap and broadening generated within proximity-coupled, ferromagnet-TI heterostructures with the mass gap and broadening created in a magnetically-doped TI,\cite{Yu2010} a material system in which the QAHE has already been experimentally demonstrated.\cite{Chang2013,Kou2014,Bestwick2015,Fox2018} In addition to the exchange splitting that is generated by the magnetic dopants, scattering of electrons on these impurities can give the states a finite lifetime that could cause a loss in quantization of the Hall conductivity. 

To quantify the broadening due to magnetic impurity scattering, we consider a random impurity potential distribution of the form
\begin{align}
	V(\v r) &= \sum_i^{N_\text{imp}} U(\v r - \v R_i), \\
	U(\v r- \v R_i) &= \left( u_x^i \sigma_x + u_y^i \sigma_y + u_z^i \sigma_z  \right) \delta (\v r - \v R_i), \label{eq:random_magnetic_impurities}
\end{align}
where $\v R_i$ are the location of the magnetic impurities, $\sigma_\alpha$ are the spin Pauli matrices,  $N_\text{imp}$ is the total number of magnetic impurities, and  $u_\alpha^i$ is the effective exchange interaction in the $\hat \alpha$ direction that is induced by the magnetic impurity. When the QAHE hall is observed experimentally,\cite{Chang2013} a small magnetic field is typically applied to align the magnetic dopants uniformly in the $\v{\hat {z}}$ direction. Therefore, we can simplify the expression for the impurity potential distribution to the case where $u_x^i = u_y^i = 0$ and $u_z^i = u_z$, allowing us to rewrite Eq.~\eqref{eq:random_magnetic_impurities} as
\begin{equation}
	U(\v r) = u_z \sigma_z  \delta (\v r). \label{eq:z_magnetic_impurities}
\end{equation}
For the next steps, it is convenient to utilize the Fourier transform of this real-space potential profile, which for the delta potential we have is given simply as
\begin{align}
	U(\v k) &= \frac{1}{\mathcal V} \int d\v r \, e^{-i \v k \cdot \v r} U(\v r)  = \frac{1}{\mathcal V} u(\v k),\\
	u(\v k) &= u_z \sigma_z.
\end{align}

In principle, this random impurity potential breaks the translational symmetry of the system, which means that momentum is no longer a good quantum number. However, an experimental sample is typically much larger than the phase coherence length in the system, so within each phase coherence length, an electron  travels under a  different random impurity potential configuration before losing its phase information. Therefore, the electron will travel under a large number of impurity configurations before being collected at a terminal, which allows us to perform an impurity self-average to restore translational invariance of the system.\cite{Bruus2004} The effect of the impurity scattering can then be calculated using an impurity-averaged self-energy, which to second order in $u_z$ is written 
\begin{align}
	\Sigma_\text{imp}(E) &= \langle V \rangle + \langle V G_0(E) V\rangle, \\
		&= \Sigma^{(1)}_\text{imp}(E) + \Sigma^{(2)}_\text{imp}(E),
\end{align}
where $\langle\rangle$ is averaging over disorder configurations and $G_0$ is the bare Green function for the massless Dirac Hamiltonian in Eq.~\eqref{eq:TI_Surface_Hamiltonian}.

We proceed by calculating the first order term of the effective self-energy, which is given simply as\cite{Bruus2004}
\begin{align}
	\Sigma^{(1)}_\text{imp}(E) &= 	\langle V \rangle = N_\text{imp} U(\v k = 0), \\
		  &= n_\text{imp} u_z \, \sigma_z \equiv M_\text{imp} \,\sigma_z	,
\end{align}
where $n_\text{imp} = \frac{N_\text{imp}}{V}$ is the density of impurities in the system. To lowest order in the exchange interaction strength, $u_z$, we see that that a random arrangement of magnetic impurities results in a net exchange splitting for the system that is proportional to the magnetic impurity density.

The second order term can be calculated in a similar fashion:\cite{Bruus2004}
\begin{align}
	\Sigma^{(2)}_\text{imp}(E) &= 	\langle V G_0(E) V\rangle \\
		&= N_\text{imp} \sum_{\v k_1} U(\v k - \v k_1) G_0(E,\v k) U(\v k_1 - \v k) \\
		\begin{split}
		&= \frac{n_\text{imp} u_z^2}{4\pi} \sigma_0 \int k \,d k \! \left[ \frac{1}{ E - \alpha k} + \frac{1 }{ E + \alpha k } \right.\\
		&\hspace{4.4em} \left.  - i \pi \delta(E - \alpha k)  - i \pi \delta(E + \alpha k)\vphantom{ \frac{1}{ E - \alpha k}}\right] 
		\end{split} 
\end{align}
(See Supplementary Material Sec.~IV for a detailed derivation). Although the real part of this expression  diverges, such an integral, can be regularized with a suitable momentum cutoff. Typical calculations choose a Brillouin zone cutoff of $\Lambda_\text{BZ} = \pi/a$, where $a$ is the lattice constant, to limit the calculation within the first Brillouin zone, so we adopt this regulator:\cite{Sakai2014}
\begin{align}
	 \text{Re}\, \Sigma^{(2)}_\text{imp}(E) &= \frac{n_\text{imp} u_z^2 }{4\pi \alpha^2} E   \ln \left( \frac{E^2}{E^2  - \frac{\alpha^2 \pi^2}{a^2}} \right)     \sigma_0 .
\end{align}
This second-order effect in $u_z$ simply raises the on-site energy and does not change the qualitative physics of the problem. Typically, we consider this contribution as a renormalization of the Fermi energy $\epsilon_F \to \epsilon_F + \lim_{E\to\epsilon_F}\text{Re}\, \Sigma^{(2)}_\text{imp}$.\cite{Groth2009}
We now focus on the imaginary part of this self-energy, which provides the electrons with a finite lifetime due to the impurity scattering:
\begin{align}
	i \text{Im}\, \Sigma^{(2)}_\text{imp}(E) &=  - i \frac{n_\text{imp} u_z^2}{4\alpha^2} |E| \sigma_0   \equiv -i \Gamma_\text{imp} \, \sigma_0.
\end{align}
This term is the leading order contribution to the broadening due to magnetic impurity scattering. Notably, it acts on both spins with the same sign and follows the exact form of the broadening due to non-magnetic impurities.\cite{Sakai2014}

Similar to the ferromagnet contact self-energy, we see that that the presence of magnetic impurities introduces both a mass gap, $M_\text{imp}$, and a non-Hermitian broadening, $\Gamma_\text{imp}$. These contributions, however, arise from different orders in the perturbation series, and it is therefore trivial to show that $\Gamma_\text{imp}/M_\text{imp} \propto u_z \ll 1$. Thus, the broadening introduced by magnetic impurities is constrained to be much smaller than the mass gap, resulting in a nearly quantized Hall conductivity based on Eqs.~\eqref{eq:sigmaxyIa_closed}-\eqref{eq:sigmaxyII_closed} and Fig.~\ref{fig:sigmaxy}. Such analysis excludes vertex corrections that naturally arise in perturbative calculations of conductivity, but the qualitative interpretation remains the same. The small broadening generated by magnetic impurity scattering is much less than that generated by a proximity-coupled ferromagnet and explains why experimental observation of the QAHE is possible in magnetic topological insulators.\cite{Chang2013}

\section{Conclusion}

In summary, we analytically derive a contact self-energy that characterizes the effect of proximity-coupling a metallic ferromagnet to a TI surface state. We show that when metallic bands from the ferromagnet are present near the Dirac point, a non-Hermitian broadening term is introduced to the Hamiltonian in addition to the anticipated time-reversal-breaking mass gap term. This contact self-energy can introduce large enough broadening that the surface spectrum appears gapless despite the presence of a mass gap. We calculate the Hall conductivity for the effective Hamiltonian describing the heterostructure and show that for any non-zero broadening, the Hall conductivity is no longer quantized. For cases when broadening is on the order of the mass gap, we find that the Hall conductivity can be significantly reduced, making the QAHE extremely challenging to observe in proximity-coupled ferromagnet-TI systems. For comparison, we also calculate the broadening and mass gap expected from scattering off of magnetic impurities in magnetically-doped TIs and find that the broadening is constrained to be much less than the mass gap. As such, the QAHE remains nearly quantized in magnetic topological insulators despite the presence of finite broadening. Therefore, our work demonstrates that although non-Hermitian Hamiltonians can be classified using non-Hermitian topological band theory, observables that are topologically quantized in the Hermitian theory, may no longer retain such quantization. 

\section{Acknowledgments}

T.M.P. and  M.J.G acknowledge financial support from the National Science Foundation (NSF) under CAREER Award ECCS-1351871. M.R.H and M.J.G acknowledge financial support from the Office of Naval Research (ONR) under grant number N00014-17-1-3012. T.M.P acknowledges fruitful discussions with Y. Kim, M. J. Park, G. A. Hamilton, and B. Basa.

\bibliography{Non-Hermitian_Topology,Magnetic_Prox_Effect}

\begin{thebibliography}{33}%
\makeatletter
\providecommand \@ifxundefined [1]{%
 \@ifx{#1\undefined}
}%
\providecommand \@ifnum [1]{%
 \ifnum #1\expandafter \@firstoftwo
 \else \expandafter \@secondoftwo
 \fi
}%
\providecommand \@ifx [1]{%
 \ifx #1\expandafter \@firstoftwo
 \else \expandafter \@secondoftwo
 \fi
}%
\providecommand \natexlab [1]{#1}%
\providecommand \enquote  [1]{``#1''}%
\providecommand \bibnamefont  [1]{#1}%
\providecommand \bibfnamefont [1]{#1}%
\providecommand \citenamefont [1]{#1}%
\providecommand \href@noop [0]{\@secondoftwo}%
\providecommand \href [0]{\begingroup \@sanitize@url \@href}%
\providecommand \@href[1]{\@@startlink{#1}\@@href}%
\providecommand \@@href[1]{\endgroup#1\@@endlink}%
\providecommand \@sanitize@url [0]{\catcode `\\12\catcode `\$12\catcode
  `\&12\catcode `\#12\catcode `\^12\catcode `\_12\catcode `\%12\relax}%
\providecommand \@@startlink[1]{}%
\providecommand \@@endlink[0]{}%
\providecommand \url  [0]{\begingroup\@sanitize@url \@url }%
\providecommand \@url [1]{\endgroup\@href {#1}{\urlprefix }}%
\providecommand \urlprefix  [0]{URL }%
\providecommand \Eprint [0]{\href }%
\providecommand \doibase [0]{http://dx.doi.org/}%
\providecommand \selectlanguage [0]{\@gobble}%
\providecommand \bibinfo  [0]{\@secondoftwo}%
\providecommand \bibfield  [0]{\@secondoftwo}%
\providecommand \translation [1]{[#1]}%
\providecommand \BibitemOpen [0]{}%
\providecommand \bibitemStop [0]{}%
\providecommand \bibitemNoStop [0]{.\EOS\space}%
\providecommand \EOS [0]{\spacefactor3000\relax}%
\providecommand \BibitemShut  [1]{\csname bibitem#1\endcsname}%
\let\auto@bib@innerbib\@empty
\bibitem [{\citenamefont {Hasan}\ and\ \citenamefont {Kane}(2010)}]{Hasan2010}%
  \BibitemOpen
  \bibfield  {author} {\bibinfo {author} {\bibfnamefont {M.~Z.}\ \bibnamefont
  {Hasan}}\ and\ \bibinfo {author} {\bibfnamefont {C.~L.}\ \bibnamefont
  {Kane}},\ }\href {\doibase 10.1103/RevModPhys.82.3045} {\bibfield  {journal}
  {\bibinfo  {journal} {Rev. Mod. Phys.}\ }\textbf {\bibinfo {volume} {82}},\
  \bibinfo {pages} {3045} (\bibinfo {year} {2010})}\BibitemShut {NoStop}%
\bibitem [{\citenamefont {Qi}\ and\ \citenamefont {Zhang}(2011)}]{Qi2011}%
  \BibitemOpen
  \bibfield  {author} {\bibinfo {author} {\bibfnamefont {X.-L.}\ \bibnamefont
  {Qi}}\ and\ \bibinfo {author} {\bibfnamefont {S.-C.}\ \bibnamefont {Zhang}},\
  }\href {\doibase 10.1103/RevModPhys.83.1057} {\bibfield  {journal} {\bibinfo
  {journal} {Rev. Mod. Phys.}\ }\textbf {\bibinfo {volume} {83}},\ \bibinfo
  {pages} {1057} (\bibinfo {year} {2011})}\BibitemShut {NoStop}%
\bibitem [{\citenamefont {Bernevig}(2013)}]{Bernevig2013}%
  \BibitemOpen
  \bibfield  {author} {\bibinfo {author} {\bibfnamefont {B.~A.}\ \bibnamefont
  {Bernevig}},\ }\href@noop {} {\emph {\bibinfo {title} {{Topological
  Insulators and Topological Superconductors}}}}\ (\bibinfo  {publisher}
  {Princeton University Press},\ \bibinfo {address} {Princeton, New Jersey},\
  \bibinfo {year} {2013})\BibitemShut {NoStop}%
\bibitem [{\citenamefont {Klitzing}\ \emph {et~al.}(1980)\citenamefont
  {Klitzing}, \citenamefont {Dorda},\ and\ \citenamefont
  {Pepper}}]{Klitzing1980a}%
  \BibitemOpen
  \bibfield  {author} {\bibinfo {author} {\bibfnamefont {K.~V.}\ \bibnamefont
  {Klitzing}}, \bibinfo {author} {\bibfnamefont {G.}~\bibnamefont {Dorda}}, \
  and\ \bibinfo {author} {\bibfnamefont {M.}~\bibnamefont {Pepper}},\ }\href
  {\doibase 10.1103/PhysRevLett.45.494} {\bibfield  {journal} {\bibinfo
  {journal} {Phys. Rev. Lett.}\ }\textbf {\bibinfo {volume} {45}},\ \bibinfo
  {pages} {494} (\bibinfo {year} {1980})},\ \Eprint
  {http://arxiv.org/abs/arXiv:1011.1669v3} {arXiv:arXiv:1011.1669v3}
  \BibitemShut {NoStop}%
\bibitem [{\citenamefont {Haldane}(1988)}]{Haldane1988}%
  \BibitemOpen
  \bibfield  {author} {\bibinfo {author} {\bibfnamefont {F.~D.~M.}\
  \bibnamefont {Haldane}},\ }\href {\doibase 10.1103/PhysRevLett.61.2015}
  {\bibfield  {journal} {\bibinfo  {journal} {Phys. Rev. Lett.}\ }\textbf
  {\bibinfo {volume} {61}},\ \bibinfo {pages} {2015} (\bibinfo {year}
  {1988})}\BibitemShut {NoStop}%
\bibitem [{\citenamefont {Jeckelmann}\ \emph {et~al.}(1997)\citenamefont
  {Jeckelmann}, \citenamefont {Jeanneret},\ and\ \citenamefont
  {Inglis}}]{Jeckelmann1997a}%
  \BibitemOpen
  \bibfield  {author} {\bibinfo {author} {\bibfnamefont {B.}~\bibnamefont
  {Jeckelmann}}, \bibinfo {author} {\bibfnamefont {B.}~\bibnamefont
  {Jeanneret}}, \ and\ \bibinfo {author} {\bibfnamefont {D.}~\bibnamefont
  {Inglis}},\ }\href {\doibase 10.1103/PhysRevB.55.13124} {\bibfield  {journal}
  {\bibinfo  {journal} {Phys. Rev. B}\ }\textbf {\bibinfo {volume} {55}},\
  \bibinfo {pages} {13124} (\bibinfo {year} {1997})}\BibitemShut {NoStop}%
\bibitem [{\citenamefont {San-Jose}\ \emph {et~al.}(2016)\citenamefont
  {San-Jose}, \citenamefont {Cayao}, \citenamefont {Prada},\ and\ \citenamefont
  {Aguado}}]{San-Jose2016}%
  \BibitemOpen
  \bibfield  {author} {\bibinfo {author} {\bibfnamefont {P.}~\bibnamefont
  {San-Jose}}, \bibinfo {author} {\bibfnamefont {J.}~\bibnamefont {Cayao}},
  \bibinfo {author} {\bibfnamefont {E.}~\bibnamefont {Prada}}, \ and\ \bibinfo
  {author} {\bibfnamefont {R.}~\bibnamefont {Aguado}},\ }\href {\doibase
  10.1038/srep21427} {\bibfield  {journal} {\bibinfo  {journal} {Sci. Rep.}\
  }\textbf {\bibinfo {volume} {6}},\ \bibinfo {pages} {21427} (\bibinfo {year}
  {2016})},\ \Eprint {http://arxiv.org/abs/1409.7306} {arXiv:1409.7306}
  \BibitemShut {NoStop}%
\bibitem [{\citenamefont {Lieu}(2018)}]{Lieu2018}%
  \BibitemOpen
  \bibfield  {author} {\bibinfo {author} {\bibfnamefont {S.}~\bibnamefont
  {Lieu}},\ }\href {\doibase 10.1103/PhysRevB.97.045106} {\bibfield  {journal}
  {\bibinfo  {journal} {Phys. Rev. B}\ }\textbf {\bibinfo {volume} {97}},\
  \bibinfo {pages} {045106} (\bibinfo {year} {2018})},\ \Eprint
  {http://arxiv.org/abs/1709.03788} {arXiv:1709.03788} \BibitemShut {NoStop}%
\bibitem [{\citenamefont {Cerjan}\ \emph {et~al.}(2018)\citenamefont {Cerjan},
  \citenamefont {Xiao}, \citenamefont {Yuan},\ and\ \citenamefont
  {Fan}}]{Cerjan2018}%
  \BibitemOpen
  \bibfield  {author} {\bibinfo {author} {\bibfnamefont {A.}~\bibnamefont
  {Cerjan}}, \bibinfo {author} {\bibfnamefont {M.}~\bibnamefont {Xiao}},
  \bibinfo {author} {\bibfnamefont {L.}~\bibnamefont {Yuan}}, \ and\ \bibinfo
  {author} {\bibfnamefont {S.}~\bibnamefont {Fan}},\ }\href {\doibase
  10.1103/PhysRevB.97.075128} {\bibfield  {journal} {\bibinfo  {journal} {Phys.
  Rev. B}\ }\textbf {\bibinfo {volume} {97}},\ \bibinfo {pages} {075128}
  (\bibinfo {year} {2018})},\ \Eprint {http://arxiv.org/abs/1712.02444}
  {arXiv:1712.02444} \BibitemShut {NoStop}%
\bibitem [{\citenamefont {Zyuzin}\ and\ \citenamefont
  {Zyuzin}(2018)}]{Zyuzin2018}%
  \BibitemOpen
  \bibfield  {author} {\bibinfo {author} {\bibfnamefont {A.~A.}\ \bibnamefont
  {Zyuzin}}\ and\ \bibinfo {author} {\bibfnamefont {A.~Y.}\ \bibnamefont
  {Zyuzin}},\ }\href {\doibase 10.1103/PhysRevB.97.041203} {\bibfield
  {journal} {\bibinfo  {journal} {Phys. Rev. B}\ }\textbf {\bibinfo {volume}
  {97}},\ \bibinfo {pages} {041203} (\bibinfo {year} {2018})},\ \Eprint
  {http://arxiv.org/abs/1710.05344} {arXiv:1710.05344} \BibitemShut {NoStop}%
\bibitem [{\citenamefont {Yao}\ \emph {et~al.}(2018)\citenamefont {Yao},
  \citenamefont {Song},\ and\ \citenamefont {Wang}}]{Yao2018}%
  \BibitemOpen
  \bibfield  {author} {\bibinfo {author} {\bibfnamefont {S.}~\bibnamefont
  {Yao}}, \bibinfo {author} {\bibfnamefont {F.}~\bibnamefont {Song}}, \ and\
  \bibinfo {author} {\bibfnamefont {Z.}~\bibnamefont {Wang}},\ }\href {\doibase
  10.1103/PhysRevLett.121.136802} {\bibfield  {journal} {\bibinfo  {journal}
  {Physical Review Letters}\ }\textbf {\bibinfo {volume} {121}},\ \bibinfo
  {pages} {136802} (\bibinfo {year} {2018})}\BibitemShut {NoStop}%
\bibitem [{\citenamefont {Zhou}\ \emph {et~al.}(2018)\citenamefont {Zhou},
  \citenamefont {Peng}, \citenamefont {Yoon}, \citenamefont {Hsu},
  \citenamefont {Nelson}, \citenamefont {Fu}, \citenamefont {Joannopoulos},
  \citenamefont {Solja{\v{c}}i{\'{c}}},\ and\ \citenamefont
  {Zhen}}]{Zhou2018a}%
  \BibitemOpen
  \bibfield  {author} {\bibinfo {author} {\bibfnamefont {H.}~\bibnamefont
  {Zhou}}, \bibinfo {author} {\bibfnamefont {C.}~\bibnamefont {Peng}}, \bibinfo
  {author} {\bibfnamefont {Y.}~\bibnamefont {Yoon}}, \bibinfo {author}
  {\bibfnamefont {C.~W.}\ \bibnamefont {Hsu}}, \bibinfo {author} {\bibfnamefont
  {K.~A.}\ \bibnamefont {Nelson}}, \bibinfo {author} {\bibfnamefont
  {L.}~\bibnamefont {Fu}}, \bibinfo {author} {\bibfnamefont {J.~D.}\
  \bibnamefont {Joannopoulos}}, \bibinfo {author} {\bibfnamefont
  {M.}~\bibnamefont {Solja{\v{c}}i{\'{c}}}}, \ and\ \bibinfo {author}
  {\bibfnamefont {B.}~\bibnamefont {Zhen}},\ }\href {\doibase
  10.1126/science.aap9859} {\bibfield  {journal} {\bibinfo  {journal}
  {Science}\ }\textbf {\bibinfo {volume} {9859}},\ \bibinfo {pages} {eaap9859}
  (\bibinfo {year} {2018})},\ \Eprint {http://arxiv.org/abs/1709.03044}
  {arXiv:1709.03044} \BibitemShut {NoStop}%
\bibitem [{\citenamefont {Shen}\ \emph {et~al.}(2018)\citenamefont {Shen},
  \citenamefont {Zhen},\ and\ \citenamefont {Fu}}]{Shen2018a}%
  \BibitemOpen
  \bibfield  {author} {\bibinfo {author} {\bibfnamefont {H.}~\bibnamefont
  {Shen}}, \bibinfo {author} {\bibfnamefont {B.}~\bibnamefont {Zhen}}, \ and\
  \bibinfo {author} {\bibfnamefont {L.}~\bibnamefont {Fu}},\ }\href {\doibase
  10.1103/PhysRevLett.120.146402} {\bibfield  {journal} {\bibinfo  {journal}
  {Phys. Rev. Lett.}\ }\textbf {\bibinfo {volume} {120}},\ \bibinfo {pages}
  {146402} (\bibinfo {year} {2018})},\ \Eprint
  {http://arxiv.org/abs/1706.07435} {arXiv:1706.07435} \BibitemShut {NoStop}%
\bibitem [{\citenamefont {Papaj}\ \emph {et~al.}(2018)\citenamefont {Papaj},
  \citenamefont {Isobe},\ and\ \citenamefont {Fu}}]{Papaj2018}%
  \BibitemOpen
  \bibfield  {author} {\bibinfo {author} {\bibfnamefont {M.}~\bibnamefont
  {Papaj}}, \bibinfo {author} {\bibfnamefont {H.}~\bibnamefont {Isobe}}, \ and\
  \bibinfo {author} {\bibfnamefont {L.}~\bibnamefont {Fu}},\ }\href
  {http://arxiv.org/abs/1802.00443} {\  (\bibinfo {year} {2018})},\ \Eprint
  {http://arxiv.org/abs/1802.00443} {arXiv:1802.00443} \BibitemShut {NoStop}%
\bibitem [{\citenamefont {Pourfath}(2014)}]{Pourfath2014}%
  \BibitemOpen
  \bibfield  {author} {\bibinfo {author} {\bibfnamefont {M.}~\bibnamefont
  {Pourfath}},\ }\href {\doibase 10.1007/978-3-7091-1800-9} {\emph {\bibinfo
  {title} {{The Non-Equilibrium Green's Function Method for Nanoscale Device
  Simulation}}}},\ Computational Microelectronics\ (\bibinfo  {publisher}
  {Springer Vienna},\ \bibinfo {address} {Vienna},\ \bibinfo {year}
  {2014})\BibitemShut {NoStop}%
\bibitem [{\citenamefont {Streda}(1982)}]{Streda1982a}%
  \BibitemOpen
  \bibfield  {author} {\bibinfo {author} {\bibfnamefont {P.}~\bibnamefont
  {Streda}},\ }\href@noop {} {\bibfield  {journal} {\bibinfo  {journal} {J.
  Phys. C Solid State Phys.}\ }\textbf {\bibinfo {volume} {15}},\ \bibinfo
  {pages} {717} (\bibinfo {year} {1982})}\BibitemShut {NoStop}%
\bibitem [{\citenamefont {Sinitsyn}\ \emph {et~al.}(2007)\citenamefont
  {Sinitsyn}, \citenamefont {MacDonald}, \citenamefont {Jungwirth},
  \citenamefont {Dugaev},\ and\ \citenamefont {Sinova}}]{Sinitsyn2007}%
  \BibitemOpen
  \bibfield  {author} {\bibinfo {author} {\bibfnamefont {N.~A.}\ \bibnamefont
  {Sinitsyn}}, \bibinfo {author} {\bibfnamefont {A.~H.}\ \bibnamefont
  {MacDonald}}, \bibinfo {author} {\bibfnamefont {T.}~\bibnamefont
  {Jungwirth}}, \bibinfo {author} {\bibfnamefont {V.~K.}\ \bibnamefont
  {Dugaev}}, \ and\ \bibinfo {author} {\bibfnamefont {J.}~\bibnamefont
  {Sinova}},\ }\href {\doibase 10.1103/PhysRevB.75.045315} {\bibfield
  {journal} {\bibinfo  {journal} {Phys. Rev. B}\ }\textbf {\bibinfo {volume}
  {75}},\ \bibinfo {pages} {045315} (\bibinfo {year} {2007})},\ \Eprint
  {http://arxiv.org/abs/0608682} {arXiv:0608682 [cond-mat]} \BibitemShut
  {NoStop}%
\bibitem [{\citenamefont {Nunner}\ \emph {et~al.}(2007)\citenamefont {Nunner},
  \citenamefont {Sinitsyn}, \citenamefont {Borunda}, \citenamefont {Dugaev},
  \citenamefont {Kovalev}, \citenamefont {Abanov}, \citenamefont {Timm},
  \citenamefont {Jungwirth}, \citenamefont {Inoue}, \citenamefont {MacDonald},\
  and\ \citenamefont {Sinova}}]{Nunner2007a}%
  \BibitemOpen
  \bibfield  {author} {\bibinfo {author} {\bibfnamefont {T.~S.}\ \bibnamefont
  {Nunner}}, \bibinfo {author} {\bibfnamefont {N.~A.}\ \bibnamefont
  {Sinitsyn}}, \bibinfo {author} {\bibfnamefont {M.~F.}\ \bibnamefont
  {Borunda}}, \bibinfo {author} {\bibfnamefont {V.~K.}\ \bibnamefont {Dugaev}},
  \bibinfo {author} {\bibfnamefont {A.~A.}\ \bibnamefont {Kovalev}}, \bibinfo
  {author} {\bibfnamefont {A.}~\bibnamefont {Abanov}}, \bibinfo {author}
  {\bibfnamefont {C.}~\bibnamefont {Timm}}, \bibinfo {author} {\bibfnamefont
  {T.}~\bibnamefont {Jungwirth}}, \bibinfo {author} {\bibfnamefont {J.-i.}\
  \bibnamefont {Inoue}}, \bibinfo {author} {\bibfnamefont {A.~H.}\ \bibnamefont
  {MacDonald}}, \ and\ \bibinfo {author} {\bibfnamefont {J.}~\bibnamefont
  {Sinova}},\ }\href {\doibase 10.1103/PhysRevB.76.235312} {\bibfield
  {journal} {\bibinfo  {journal} {Phys. Rev. B}\ }\textbf {\bibinfo {volume}
  {76}},\ \bibinfo {pages} {235312} (\bibinfo {year} {2007})},\ \Eprint
  {http://arxiv.org/abs/0502386} {arXiv:0502386 [cond-mat]} \BibitemShut
  {NoStop}%
\bibitem [{\citenamefont {Nagaosa}\ \emph {et~al.}(2010)\citenamefont
  {Nagaosa}, \citenamefont {Sinova}, \citenamefont {Onoda}, \citenamefont
  {MacDonald},\ and\ \citenamefont {Ong}}]{Nagaosa2010a}%
  \BibitemOpen
  \bibfield  {author} {\bibinfo {author} {\bibfnamefont {N.}~\bibnamefont
  {Nagaosa}}, \bibinfo {author} {\bibfnamefont {J.}~\bibnamefont {Sinova}},
  \bibinfo {author} {\bibfnamefont {S.}~\bibnamefont {Onoda}}, \bibinfo
  {author} {\bibfnamefont {A.~H.}\ \bibnamefont {MacDonald}}, \ and\ \bibinfo
  {author} {\bibfnamefont {N.~P.}\ \bibnamefont {Ong}},\ }\href {\doibase
  10.1103/RevModPhys.82.1539} {\bibfield  {journal} {\bibinfo  {journal} {Rev.
  Mod. Phys.}\ }\textbf {\bibinfo {volume} {82}},\ \bibinfo {pages} {1539}
  (\bibinfo {year} {2010})}\BibitemShut {NoStop}%
\bibitem [{\citenamefont {Batallan}\ \emph {et~al.}(1975)\citenamefont
  {Batallan}, \citenamefont {Rosenman},\ and\ \citenamefont
  {Sommers}}]{Batallan1975}%
  \BibitemOpen
  \bibfield  {author} {\bibinfo {author} {\bibfnamefont {F.}~\bibnamefont
  {Batallan}}, \bibinfo {author} {\bibfnamefont {I.}~\bibnamefont {Rosenman}},
  \ and\ \bibinfo {author} {\bibfnamefont {C.~B.}\ \bibnamefont {Sommers}},\
  }\href {\doibase 10.1103/PhysRevB.11.545} {\bibfield  {journal} {\bibinfo
  {journal} {Physical Review B}\ }\textbf {\bibinfo {volume} {11}},\ \bibinfo
  {pages} {545} (\bibinfo {year} {1975})}\BibitemShut {NoStop}%
\bibitem [{\citenamefont {Barreteau}\ \emph {et~al.}(2016)\citenamefont
  {Barreteau}, \citenamefont {Spanjaard},\ and\ \citenamefont
  {Desjonqu{\`{e}}res}}]{Barreteau2016}%
  \BibitemOpen
  \bibfield  {author} {\bibinfo {author} {\bibfnamefont {C.}~\bibnamefont
  {Barreteau}}, \bibinfo {author} {\bibfnamefont {D.}~\bibnamefont
  {Spanjaard}}, \ and\ \bibinfo {author} {\bibfnamefont {M.-C.}\ \bibnamefont
  {Desjonqu{\`{e}}res}},\ }\href {\doibase 10.1016/j.crhy.2015.12.014}
  {\bibfield  {journal} {\bibinfo  {journal} {Comptes Rendus Physique}\
  }\textbf {\bibinfo {volume} {17}},\ \bibinfo {pages} {406} (\bibinfo {year}
  {2016})}\BibitemShut {NoStop}%
\bibitem [{\citenamefont {Cullity}\ and\ \citenamefont
  {Graham}(2009)}]{Cullity2009}%
  \BibitemOpen
  \bibfield  {author} {\bibinfo {author} {\bibfnamefont {B.~D.}\ \bibnamefont
  {Cullity}}\ and\ \bibinfo {author} {\bibfnamefont {C.~D.}\ \bibnamefont
  {Graham}},\ }\href@noop {} {\emph {\bibinfo {title} {{Introduction to
  magnetic materials}}}},\ \bibinfo {edition} {2nd}\ ed.\ (\bibinfo
  {publisher} {John Wiley {\&} Sons},\ \bibinfo {address} {New Jersey},\
  \bibinfo {year} {2009})\BibitemShut {NoStop}%
\bibitem [{\citenamefont {Luo}\ and\ \citenamefont {Qi}(2013)}]{Luo2013a}%
  \BibitemOpen
  \bibfield  {author} {\bibinfo {author} {\bibfnamefont {W.}~\bibnamefont
  {Luo}}\ and\ \bibinfo {author} {\bibfnamefont {X.-L.}\ \bibnamefont {Qi}},\
  }\href {\doibase 10.1103/PhysRevB.87.085431} {\bibfield  {journal} {\bibinfo
  {journal} {Phys. Rev. B}\ }\textbf {\bibinfo {volume} {87}},\ \bibinfo
  {pages} {085431} (\bibinfo {year} {2013})},\ \Eprint
  {http://arxiv.org/abs/1208.4638} {arXiv:1208.4638} \BibitemShut {NoStop}%
\bibitem [{\citenamefont {Bahramy}\ \emph {et~al.}(2012)\citenamefont
  {Bahramy}, \citenamefont {King}, \citenamefont {de~la Torre}, \citenamefont
  {Chang}, \citenamefont {Shi}, \citenamefont {Patthey}, \citenamefont
  {Balakrishnan}, \citenamefont {Hofmann}, \citenamefont {Arita}, \citenamefont
  {Nagaosa},\ and\ \citenamefont {Baumberger}}]{Bahramy2012}%
  \BibitemOpen
  \bibfield  {author} {\bibinfo {author} {\bibfnamefont {M.}~\bibnamefont
  {Bahramy}}, \bibinfo {author} {\bibfnamefont {P.}~\bibnamefont {King}},
  \bibinfo {author} {\bibfnamefont {A.}~\bibnamefont {de~la Torre}}, \bibinfo
  {author} {\bibfnamefont {J.}~\bibnamefont {Chang}}, \bibinfo {author}
  {\bibfnamefont {M.}~\bibnamefont {Shi}}, \bibinfo {author} {\bibfnamefont
  {L.}~\bibnamefont {Patthey}}, \bibinfo {author} {\bibfnamefont
  {G.}~\bibnamefont {Balakrishnan}}, \bibinfo {author} {\bibfnamefont
  {P.}~\bibnamefont {Hofmann}}, \bibinfo {author} {\bibfnamefont
  {R.}~\bibnamefont {Arita}}, \bibinfo {author} {\bibfnamefont
  {N.}~\bibnamefont {Nagaosa}}, \ and\ \bibinfo {author} {\bibfnamefont
  {F.}~\bibnamefont {Baumberger}},\ }\href {\doibase 10.1038/ncomms2162}
  {\bibfield  {journal} {\bibinfo  {journal} {Nat. Commun.}\ }\textbf {\bibinfo
  {volume} {3}},\ \bibinfo {pages} {1159} (\bibinfo {year} {2012})},\ \Eprint
  {http://arxiv.org/abs/1206.0564} {arXiv:1206.0564} \BibitemShut {NoStop}%
\bibitem [{\citenamefont {Garate}\ and\ \citenamefont
  {Glazman}(2012)}]{Garate2012}%
  \BibitemOpen
  \bibfield  {author} {\bibinfo {author} {\bibfnamefont {I.}~\bibnamefont
  {Garate}}\ and\ \bibinfo {author} {\bibfnamefont {L.}~\bibnamefont
  {Glazman}},\ }\href {\doibase 10.1103/PhysRevB.86.035422} {\bibfield
  {journal} {\bibinfo  {journal} {Phys. Rev. B}\ }\textbf {\bibinfo {volume}
  {86}},\ \bibinfo {pages} {035422} (\bibinfo {year} {2012})},\ \Eprint
  {http://arxiv.org/abs/1206.1239} {arXiv:1206.1239} \BibitemShut {NoStop}%
\bibitem [{\citenamefont {Yu}\ \emph {et~al.}(2010)\citenamefont {Yu},
  \citenamefont {Zhang}, \citenamefont {Zhang}, \citenamefont {Zhang},
  \citenamefont {Dai},\ and\ \citenamefont {Fang}}]{Yu2010}%
  \BibitemOpen
  \bibfield  {author} {\bibinfo {author} {\bibfnamefont {R.}~\bibnamefont
  {Yu}}, \bibinfo {author} {\bibfnamefont {W.}~\bibnamefont {Zhang}}, \bibinfo
  {author} {\bibfnamefont {H.-J.}\ \bibnamefont {Zhang}}, \bibinfo {author}
  {\bibfnamefont {S.-C.}\ \bibnamefont {Zhang}}, \bibinfo {author}
  {\bibfnamefont {X.}~\bibnamefont {Dai}}, \ and\ \bibinfo {author}
  {\bibfnamefont {Z.}~\bibnamefont {Fang}},\ }\href {\doibase
  10.1126/science.1187485} {\bibfield  {journal} {\bibinfo  {journal}
  {Science}\ }\textbf {\bibinfo {volume} {329}},\ \bibinfo {pages} {61}
  (\bibinfo {year} {2010})}\BibitemShut {NoStop}%
\bibitem [{\citenamefont {Chang}\ \emph {et~al.}(2013)\citenamefont {Chang},
  \citenamefont {Zhang}, \citenamefont {Feng}, \citenamefont {Shen},
  \citenamefont {Zhang}, \citenamefont {Guo}, \citenamefont {Li}, \citenamefont
  {Ou}, \citenamefont {Wei}, \citenamefont {Wang}, \citenamefont {Ji},
  \citenamefont {Feng}, \citenamefont {Ji}, \citenamefont {Chen}, \citenamefont
  {Jia}, \citenamefont {Dai}, \citenamefont {Fang}, \citenamefont {Zhang},
  \citenamefont {He}, \citenamefont {Wang}, \citenamefont {Lu}, \citenamefont
  {Ma},\ and\ \citenamefont {Xue}}]{Chang2013}%
  \BibitemOpen
  \bibfield  {author} {\bibinfo {author} {\bibfnamefont {C.}~\bibnamefont
  {Chang}}, \bibinfo {author} {\bibfnamefont {J.}~\bibnamefont {Zhang}},
  \bibinfo {author} {\bibfnamefont {X.}~\bibnamefont {Feng}}, \bibinfo {author}
  {\bibfnamefont {J.}~\bibnamefont {Shen}}, \bibinfo {author} {\bibfnamefont
  {Z.}~\bibnamefont {Zhang}}, \bibinfo {author} {\bibfnamefont
  {M.}~\bibnamefont {Guo}}, \bibinfo {author} {\bibfnamefont {K.}~\bibnamefont
  {Li}}, \bibinfo {author} {\bibfnamefont {Y.}~\bibnamefont {Ou}}, \bibinfo
  {author} {\bibfnamefont {P.}~\bibnamefont {Wei}}, \bibinfo {author}
  {\bibfnamefont {L.-L.}\ \bibnamefont {Wang}}, \bibinfo {author}
  {\bibfnamefont {Z.-Q.}\ \bibnamefont {Ji}}, \bibinfo {author} {\bibfnamefont
  {Y.}~\bibnamefont {Feng}}, \bibinfo {author} {\bibfnamefont {S.}~\bibnamefont
  {Ji}}, \bibinfo {author} {\bibfnamefont {X.}~\bibnamefont {Chen}}, \bibinfo
  {author} {\bibfnamefont {J.}~\bibnamefont {Jia}}, \bibinfo {author}
  {\bibfnamefont {X.}~\bibnamefont {Dai}}, \bibinfo {author} {\bibfnamefont
  {Z.}~\bibnamefont {Fang}}, \bibinfo {author} {\bibfnamefont {S.-C.}\
  \bibnamefont {Zhang}}, \bibinfo {author} {\bibfnamefont {K.}~\bibnamefont
  {He}}, \bibinfo {author} {\bibfnamefont {Y.}~\bibnamefont {Wang}}, \bibinfo
  {author} {\bibfnamefont {L.}~\bibnamefont {Lu}}, \bibinfo {author}
  {\bibfnamefont {X.-C.}\ \bibnamefont {Ma}}, \ and\ \bibinfo {author}
  {\bibfnamefont {Q.-K.}\ \bibnamefont {Xue}},\ }\href {\doibase
  10.1126/science.1234414} {\bibfield  {journal} {\bibinfo  {journal}
  {Science}\ }\textbf {\bibinfo {volume} {340}},\ \bibinfo {pages} {167}
  (\bibinfo {year} {2013})}\BibitemShut {NoStop}%
\bibitem [{\citenamefont {Kou}\ \emph {et~al.}(2014)\citenamefont {Kou},
  \citenamefont {Guo}, \citenamefont {Fan}, \citenamefont {Pan}, \citenamefont
  {Lang}, \citenamefont {Jiang}, \citenamefont {Shao}, \citenamefont {Nie},
  \citenamefont {Murata}, \citenamefont {Tang}, \citenamefont {Wang},
  \citenamefont {He}, \citenamefont {Lee}, \citenamefont {Lee},\ and\
  \citenamefont {Wang}}]{Kou2014}%
  \BibitemOpen
  \bibfield  {author} {\bibinfo {author} {\bibfnamefont {X.}~\bibnamefont
  {Kou}}, \bibinfo {author} {\bibfnamefont {S.-T.}\ \bibnamefont {Guo}},
  \bibinfo {author} {\bibfnamefont {Y.}~\bibnamefont {Fan}}, \bibinfo {author}
  {\bibfnamefont {L.}~\bibnamefont {Pan}}, \bibinfo {author} {\bibfnamefont
  {M.}~\bibnamefont {Lang}}, \bibinfo {author} {\bibfnamefont {Y.}~\bibnamefont
  {Jiang}}, \bibinfo {author} {\bibfnamefont {Q.}~\bibnamefont {Shao}},
  \bibinfo {author} {\bibfnamefont {T.}~\bibnamefont {Nie}}, \bibinfo {author}
  {\bibfnamefont {K.}~\bibnamefont {Murata}}, \bibinfo {author} {\bibfnamefont
  {J.}~\bibnamefont {Tang}}, \bibinfo {author} {\bibfnamefont {Y.}~\bibnamefont
  {Wang}}, \bibinfo {author} {\bibfnamefont {L.}~\bibnamefont {He}}, \bibinfo
  {author} {\bibfnamefont {T.-K.}\ \bibnamefont {Lee}}, \bibinfo {author}
  {\bibfnamefont {W.-L.}\ \bibnamefont {Lee}}, \ and\ \bibinfo {author}
  {\bibfnamefont {K.~L.}\ \bibnamefont {Wang}},\ }\href {\doibase
  10.1103/PhysRevLett.113.137201} {\bibfield  {journal} {\bibinfo  {journal}
  {Physical Review Letters}\ }\textbf {\bibinfo {volume} {113}},\ \bibinfo
  {pages} {137201} (\bibinfo {year} {2014})}\BibitemShut {NoStop}%
\bibitem [{\citenamefont {Bestwick}\ \emph {et~al.}(2015)\citenamefont
  {Bestwick}, \citenamefont {Fox}, \citenamefont {Kou}, \citenamefont {Pan},
  \citenamefont {Wang},\ and\ \citenamefont {Goldhaber-Gordon}}]{Bestwick2015}%
  \BibitemOpen
  \bibfield  {author} {\bibinfo {author} {\bibfnamefont {A.~J.}\ \bibnamefont
  {Bestwick}}, \bibinfo {author} {\bibfnamefont {E.~J.}\ \bibnamefont {Fox}},
  \bibinfo {author} {\bibfnamefont {X.}~\bibnamefont {Kou}}, \bibinfo {author}
  {\bibfnamefont {L.}~\bibnamefont {Pan}}, \bibinfo {author} {\bibfnamefont
  {K.~L.}\ \bibnamefont {Wang}}, \ and\ \bibinfo {author} {\bibfnamefont
  {D.}~\bibnamefont {Goldhaber-Gordon}},\ }\href {\doibase
  10.1103/PhysRevLett.114.187201} {\bibfield  {journal} {\bibinfo  {journal}
  {Physical Review Letters}\ }\textbf {\bibinfo {volume} {114}},\ \bibinfo
  {pages} {187201} (\bibinfo {year} {2015})},\ \Eprint
  {http://arxiv.org/abs/1412.3189} {arXiv:1412.3189} \BibitemShut {NoStop}%
\bibitem [{\citenamefont {Fox}\ \emph {et~al.}(2018)\citenamefont {Fox},
  \citenamefont {Rosen}, \citenamefont {Yang}, \citenamefont {Jones},
  \citenamefont {Elmquist}, \citenamefont {Kou}, \citenamefont {Pan},
  \citenamefont {Wang},\ and\ \citenamefont {Goldhaber-Gordon}}]{Fox2018}%
  \BibitemOpen
  \bibfield  {author} {\bibinfo {author} {\bibfnamefont {E.~J.}\ \bibnamefont
  {Fox}}, \bibinfo {author} {\bibfnamefont {I.~T.}\ \bibnamefont {Rosen}},
  \bibinfo {author} {\bibfnamefont {Y.}~\bibnamefont {Yang}}, \bibinfo {author}
  {\bibfnamefont {G.~R.}\ \bibnamefont {Jones}}, \bibinfo {author}
  {\bibfnamefont {R.~E.}\ \bibnamefont {Elmquist}}, \bibinfo {author}
  {\bibfnamefont {X.}~\bibnamefont {Kou}}, \bibinfo {author} {\bibfnamefont
  {L.}~\bibnamefont {Pan}}, \bibinfo {author} {\bibfnamefont {K.~L.}\
  \bibnamefont {Wang}}, \ and\ \bibinfo {author} {\bibfnamefont
  {D.}~\bibnamefont {Goldhaber-Gordon}},\ }\href {\doibase
  10.1103/PhysRevB.98.075145} {\bibfield  {journal} {\bibinfo  {journal}
  {Physical Review B}\ }\textbf {\bibinfo {volume} {98}},\ \bibinfo {pages}
  {075145} (\bibinfo {year} {2018})},\ \Eprint
  {http://arxiv.org/abs/1710.01850} {arXiv:1710.01850} \BibitemShut {NoStop}%
\bibitem [{\citenamefont {Bruus}\ and\ \citenamefont
  {Flensberg}(2004)}]{Bruus2004}%
  \BibitemOpen
  \bibfield  {author} {\bibinfo {author} {\bibfnamefont {H.}~\bibnamefont
  {Bruus}}\ and\ \bibinfo {author} {\bibfnamefont {K.}~\bibnamefont
  {Flensberg}},\ }\href
  {http://stacks.iop.org/0305-4470/38/i=8/a=B01?key=crossref.f643726123fec03a4ccc05cd9639f5cf}
  {\emph {\bibinfo {title} {{Many-Body Quantum Theory in Condensed Matter
  Physics: An Introduction}}}}\ (\bibinfo  {publisher} {Oxford University
  Press},\ \bibinfo {address} {Oxford},\ \bibinfo {year} {2004})\BibitemShut
  {NoStop}%
\bibitem [{\citenamefont {Sakai}\ and\ \citenamefont
  {Kohno}(2014)}]{Sakai2014}%
  \BibitemOpen
  \bibfield  {author} {\bibinfo {author} {\bibfnamefont {A.}~\bibnamefont
  {Sakai}}\ and\ \bibinfo {author} {\bibfnamefont {H.}~\bibnamefont {Kohno}},\
  }\href {\doibase 10.1103/PhysRevB.89.165307} {\bibfield  {journal} {\bibinfo
  {journal} {Phys. Rev. B}\ }\textbf {\bibinfo {volume} {89}},\ \bibinfo
  {pages} {165307} (\bibinfo {year} {2014})},\ \Eprint
  {http://arxiv.org/abs/arXiv:1309.4195v1} {arXiv:arXiv:1309.4195v1}
  \BibitemShut {NoStop}%
\bibitem [{\citenamefont {Groth}\ \emph {et~al.}(2009)\citenamefont {Groth},
  \citenamefont {Wimmer}, \citenamefont {Akhmerov}, \citenamefont
  {Tworzyd{\l}o},\ and\ \citenamefont {Beenakker}}]{Groth2009}%
  \BibitemOpen
  \bibfield  {author} {\bibinfo {author} {\bibfnamefont {C.~W.}\ \bibnamefont
  {Groth}}, \bibinfo {author} {\bibfnamefont {M.}~\bibnamefont {Wimmer}},
  \bibinfo {author} {\bibfnamefont {A.~R.}\ \bibnamefont {Akhmerov}}, \bibinfo
  {author} {\bibfnamefont {J.}~\bibnamefont {Tworzyd{\l}o}}, \ and\ \bibinfo
  {author} {\bibfnamefont {C.~W.}\ \bibnamefont {Beenakker}},\ }\href {\doibase
  10.1103/PhysRevLett.103.196805} {\bibfield  {journal} {\bibinfo  {journal}
  {Phys. Rev. Lett.}\ }\textbf {\bibinfo {volume} {103}},\ \bibinfo {pages} {1}
  (\bibinfo {year} {2009})},\ \Eprint {http://arxiv.org/abs/0908.0881}
  {arXiv:0908.0881} \BibitemShut {NoStop}%
\end{thebibliography}%


%

\end{document}